\documentclass[aps,prd,showpacs,nobibnotes,nofootinbib,twocolumn,preprintnumbers,nobalancelastpage,superscriptaddress]{revtex4}
\usepackage{graphicx}
\usepackage{latexsym}
\usepackage{dcolumn}
\usepackage{bm}
\input{epsf}
\newcommand{\be}{\begin{equation}}
\newcommand{\ee}{\end{equation}}
\newcommand{\ba}{\begin{eqnarray}}
\newcommand{\ea}{\end{eqnarray}}

\newcommand{\bfi}{\begin{figure}
\epsfxsize=9cm
\epsffile}
\newcommand{\efi}{\end{figure}}
\newcommand{\bi}{\begin{itemize}}
\newcommand{\ei}{\end{itemize}}

\newcommand{\app}{Astropart. Phys. }

\newcommand{\nima}{Nucl. Instrum. Methods Phys. Res., Sect. A }
\newcommand{\etal}{\emph{et al.}}

\usepackage{epsfig}

\begin{document}

\date{\today}
\title{Measurement of the cosmic ray antiproton/proton flux ratio at TeV energies with the ARGO-YBJ detector.}

\author{B.~Bartoli}
 \affiliation{Dipartimento di Fisica dell'Universit\`a di Napoli
                  ``Federico II'', Complesso Universitario di Monte 
                  Sant'Angelo, via Cinthia, 80126 Napoli, Italy.}
 \affiliation{Istituto Nazionale di Fisica Nucleare, Sezione di
                  Napoli, Complesso Universitario di Monte
                  Sant'Angelo, via Cinthia, 80126 Napoli, Italy.}
\author{P.~Bernardini}
 \affiliation{Dipartimento di Fisica dell'Universit\`a del Salento,
                  via per Arnesano, 73100 Lecce, Italy.}
 \affiliation{Istituto Nazionale di Fisica Nucleare, Sezione di
                  Lecce, via per Arnesano, 73100 Lecce, Italy.}
\author{X.J.~Bi}
 \affiliation{Key Laboratory of Particle Astrophysics, Institute 
                  of High Energy Physics, Chinese Academy of Sciences,
                  P.O. Box 918, 100049 Beijing, P.R. China.}
\author{C.~Bleve}
 \affiliation{Dipartimento di Fisica dell'Universit\`a del Salento,
                  via per Arnesano, 73100 Lecce, Italy.}
 \affiliation{Istituto Nazionale di Fisica Nucleare, Sezione di
                  Lecce, via per Arnesano, 73100 Lecce, Italy.}
\author{I.~Bolognino}
 \affiliation{Dipartimento di Fisica Nucleare e Teorica 
                  dell'Universit\`a di Pavia, via Bassi 6,
                  27100 Pavia, Italy.}
 \affiliation{Istituto Nazionale di Fisica Nucleare, Sezione di Pavia, 
                  via Bassi 6, 27100 Pavia, Italy.}
\author{P.~Branchini}
  \affiliation{Istituto Nazionale di Fisica Nucleare, Sezione di
                  Roma Tre, via della Vasca Navale 84, 00146 Roma, Italy.}
\author{A.~Budano}
 \affiliation{Istituto Nazionale di Fisica Nucleare, Sezione di
                  Roma Tre, via della Vasca Navale 84, 00146 Roma, Italy.}
\author{A.K.~Calabrese Melcarne}
 \affiliation{Istituto Nazionale di Fisica Nucleare - CNAF, Viale 
                  Berti-Pichat 6/2, 40127 Bologna, Italy.}
\author{P.~Camarri}
 \affiliation{Dipartimento di Fisica dell'Universit\`a di Roma ``Tor 									   Vergata'', via della Ricerca Scientifica 1, 00133 Roma, Italy.}
 \affiliation{Istituto Nazionale di Fisica Nucleare, Sezione di
                   Roma Tor Vergata, via della Ricerca Scientifica 1, 
                   00133 Roma, Italy.}
\author{Z.~Cao}
 \affiliation{Key Laboratory of Particle Astrophysics, Institute 
                  of High Energy Physics, Chinese Academy of Sciences,
                  P.O. Box 918, 100049 Beijing, P.R. China.}
% ...

\collaboration{ARGO-YBJ Collaboration}

%

\begin{abstract}
Cosmic ray antiprotons provide an important probe to study the
cosmic ray propagation in the interstellar space and to
investigate the existence of dark matter. 
Acting the Earth-Moon system as a magnetic spectrometer, paths of primary antiprotons are deflected in the opposite sense with respect to those of the protons in their way to the Earth.
This effect allows, in principle, the search for antiparticles in the direction opposite to the observed deficit of cosmic rays due to the Moon (the so-called \emph{`Moon shadow'}).

The ARGO-YBJ experiment, located at the Yangbajing Cosmic Ray
Laboratory (Tibet, P.R. China, 4300 m a.s.l., 606 g/cm$^2$), is particularly effective in measuring the cosmic ray antimatter content via the observation of the cosmic rays shadowing effect due to: (1) good angular resolution, pointing accuracy and long-term stability; (2) low energy threshold; (3) real sensitivity to the geomagnetic field.

Based on all the data recorded during the period from July 2006 through November 2009 and on a full Monte Carlo simulation, we searched for the existence of the shadow cast by antiprotons in the TeV energy region. No evidence of the existence of antiprotons is found in this energy region.
Upper limits to the $\bar{p}/p$ flux ratio are set to 5\% at a median energy of 1.4 TeV and 6\% at 5 TeV with a confidence level of 90\%. In the TeV energy range these limits are the lowest available.

\end{abstract}
\maketitle

\section{Introduction}

\subsection{Cosmic ray antiproton production}

Very high energy cosmic ray (VHE CR) antiprotons are an essential diagnostic tool to approach the solution of several important questions of cosmology, astrophysics and particle physics, besides studying fundamental properties of the CR sources and propagation medium. The enigma of the matter/antimatter asymmetry in the local Universe, that of the existence of antimatter regions, the search for signatures of physics beyond the standard model of particles and fields, as well as the determination of the essential features of CR propagation in the insterstellar medium, these are only a few research topics that would greatly benefit from the detection of VHE antiprotons (see for example \cite{stecker02,dine04}). 


First of all, the observation of $\overline{p}$ abundance in the CR
flux is a key to understand CR propagation. In fact, antiprotons are produced by standard nuclear interactions of CR nuclei with the interstellar medium, the information coming from these spallation processes being complementary to that achievable from secondary nuclei like Li, Be and B or secondaries of iron. It should be noticed that antiprotons mostly trace the propagation history of protons \emph{pp$\rightarrow$ $\overline{p}$ ppp}, unlike the other spallation products, which may come from heavier nuclei. Also secondary $\overline{p}$ represent a background flux that must be carefully determined to take out a primary $\overline{p}$ component due to any hypothetical exotic signal.

The observed amount of antiprotons in CRs is still far from being figured out. Detailed calculations show that there is no model capable of
accurately describing altogether B/C and sub-Fe/Fe ratios, spectra of p, He,
$\overline{p}$, e$^+$, e$^-$ and diffuse $\gamma$-rays. In fact, conventional models without reacceleration fail in reproducing both B/C ratio and $\overline{p}$ flux at the same time \cite{moskalenko02}. Diffusive reacceleration models naturally reproduce secondary/primary nuclei ratio in CRs but produce too few antiprotons \cite{moskalenko03}. The introduction of a break in the diffusion coefficient \cite{moskalenko02} would lead to consistent results, but it is not theoretically justified so far, still resulting as an \emph{ad hoc} assumption \cite{moskalenko03}.
Some models taking into account Galactic convective wind and stochastic reacceleration may reproduce both antiproton flux and secondary/primary ratio \cite{donato01}.

The estimates of the secondary $\overline{p}$ flux are suffering from uncertainties on models and parameters of particle
propagation in the Galaxy, CR spectrum and composition, details of
the nuclear cross sections for $\overline{p}$ production,
annihilation and scattering and, finally, on the heliospheric
modulation \cite{strong07,donato01}. On the contrary, there is a general
agreement in the calculation of the secondary high-energy
$\overline{p}$ flux falloff: around 50 GeV the intensity decreases
by about 3 orders of magnitude below the maximum.

Recent measurements of the antiproton flux up to about 180 GeV
by the PAMELA satellite \cite{pamelap,pamelap2} are consistent with the conventional CR model, in which antiprotons are secondary particles yielded by the spallation of CR nuclei over the interstellar medium.
Nevertheless, given the current uncertainties on propagation parameters, exotic models of primary $\overline{p}$ production cannot be ruled out \cite{pamelap2,wino09}. 
As an example, recent calculations suggest that the overall PAMELA  $\overline{p}$ and $e^+$ data \cite{pamelap2,pamelae} and Fermi 
$e^++e^-$ data \cite{fermielect} can be reproduced taking into account a heavy dark matter particle (M $\geq$10 TeV) that annihilates into $W^+W^-$ or $hh$ \cite{cirelli}. 
This scenario implies that the $\overline{p}/p$ ratio, consistent with
the background of secondary production up to about 50 GeV,
increases rapidly reaching the 10$^{-2}$ level at about 2 TeV.
CR antiprotons, as well as positrons, are therefore considered as prime targets for indirect detection of dark matter \cite{dm,donato09,evoli11}.

But it has been also suggested that the PAMELA positron data may be a natural consequence of the standard scenario for the origin of galactic CRs, if secondary e$^+$ (and e$^-$) production takes place in the same region where CRs are being accelerated \cite{blasi1}. Since this is a hadronic mechanism, an associated rise of the $\overline{p}/p$ ratio is predicted at energies $\geq$ 1 TeV \cite{blasip}.
Therefore, the high-energy range of the antiproton spectrum may reveal important constraints on the physics of the CR acceleration sites.

Antiprotons can be produced from primordial black holes
evaporation \cite{pbh1,pbh2} or in antigalaxies \cite{golden,stecker84,stecker85,stephens85}. 
In particular, it seems possible that mechanisms exist that could produce the formation of separated antimatter domains during the cosmic evolution, thus allowing the visible Universe to be globally matter-antimatter symmetric \cite{klopov00}.
In addition, the possibility exists to have antimatter confined into condensed bodies like antistars in our Galaxy \cite{duda94,klopov98}.

In theories in which matter and antimatter are present in equal
amounts in spatially separated domains of survivable size it is
expected that the $\overline{p}/p$ ratio should increase with
energy in the framework of the energy-dependent confinement model
for CRs in the Galaxy. 
In a simple "leaky box" model the energy spectrum is solely
determined by the balance between generation at the source and
escape from the Galaxy. If the source spectrum is proportional to
E$^{-\alpha}$, the equilibrium spectrum of CRs inside the source
region (i.e. inside the Galaxy) would be $\propto E^{-(\alpha+\delta)}$, due to the energy-dependent leakage of the source. 
Indeed, according to recent measurements of B/C ratio in the primary
CRs up to about 50 GeV/nucleon, the residence time of CRs in the Galaxy
can be described by a power law in energy or rigidity $\propto
R^{-\delta}$, where $\delta\sim$0.6 \cite{strong07}.
As a result, the energy spectrum of the CRs leaked
out from the antigalaxy has a spectrum $\propto E^{-\alpha}$.
If we assume that there exists a general acceleration mechanism for generating CRs which acts in both galactic and extragalactic sources to give an universal source spectrum E$^{-\alpha}$, the extragalactic CR spectrum should reflect it. 
Thus, if antiprotons are assumed to be both primary and extragalactic, we should observe the source spectrum of antigalaxies $\propto E^{-\alpha}$ and the expected $\overline{p}/p$ ratio should increase with energy:
$\overline{p}/p \propto E^{\delta}$ \cite{stecker84,stecker85,}.
As a consequence, the antiproton fraction could increase up to about 1\% around 500 GeV or even to 50\% in the multi-TeV energy range with important observational implications, being, at these energies, the background of secondary antiprotons well below this prediction. 

At high energies ($\geq$ 100 GeV) the main observable related to the
residence time of CRs in the Galaxy is the large-scale anisotropy
in their arrival direction that is known to be strictly related to the diffusion coefficient. Measurements give the amplitude of
the first angular harmonic of anisotropy at the order 10$^{-3}$ in the energy range 10$^{11}$ to 10$^{14}$ eV where the most reliable data are available \cite{guillian07,tibet06}. The data on CR anisotropy are consistent, within a factor of about 3, with a diffusion coefficient increasing with energy $\propto E^{0.3}$, as predicted in models including stochastic reacceleration by Kolmogorov-type hydromagnetic turbolence (2nd order Fermi acceleration) \cite{strong07}. 
The measurement of the $\overline{p}/p$ ratio at high energies may be useful to constrain models for $\overline{p}$ production and for the confinement of CRs, even if it is not straightforward to infer the propagation parameters, as the diffusion index $\delta$, since they are partially degenerate with source parameters \cite{blasip}.

The first antiproton upper limits in the high energy region have
been obtained by Stephens in 1985 exploiting the observed charge
ratio of muons at sea level. The presence of $\overline{p}$
dilutes the charge ratio $\mu^+/\mu^-$. The limits thus derived
for the $\overline{p}/p$ ratio are 7\%, 17\%, 10\% and 14\%
respectively for the energy intervals 0.1 - 0.2 TeV, 1.0 - 1.5
TeV, 10 - 15 TeV and $>$30 TeV \cite{stephens85}.

In addition, deeper measurements of the $\overline{p}/p$ ratio at
high energies have been performed exploiting the Earth-Moon system
as a magnetic spectrometer able to disentangle, in principle, the
deflection of protons from that of antiprotons in the geomagnetic
field.

\subsection{The Moon shadowing effect}

Since the Moon has an angular radius of about 0.26$^{\circ}$, it
must cast a shadow in the nearly-isotropic CR flux (the
so-called \emph{shadow of the Moon}). As first suggested by Clark
in 1957 \cite{clark}, the shadowing of CRs from the
direction of the Moon is useful in measuring the angular
resolution of an air shower array directly, without needing Monte Carlo (MC) simulations. In fact, the shape of the shadow provides a measurement of the detector point spread function, and its position allows the check of possible pointing biases.

In addition, due to the geomagnetic field (GMF), positively charged
particles are deflected by an angle depending on the primary CR energy \cite{argo-moon11}. This effect produces a displacement of the shadow towards the West with respect to the Moon position and smears the shape in the East-West direction, especially at low energies. The observation of the displacement of the Moon provides a direct calibration of the relation between shower size and primary energy \cite{argo-moon11}.

The same shadowing effect can be seen in the direction of the Sun.
Nevertheless, the interpretation of the shadow phenomenology is
more complex. In fact, the displacement of the shadow from the
apparent position of the Sun could be explained by the joint
effects of the GMF and of the Solar and Interplanetary Magnetic Fields (SMF and IMF, respectively) whose configuration
considerably changes with the phases of the solar activity cycle
\cite{amenomori-sun00,argo-sun11}.

Linsley \cite{linsley} and Lloyd-Evans \cite{lloyd} in 1985,
following a Watson's suggestion, independently explored the
possibility to use the Moon or Sun shadows as mass spectrometers
in order to measure the charge composition of CR spectrum.
In particular Linsley first discussed the idea to measure the
CR antiprotons abundance exploiting the separation of the
proton and antiproton shadows. In 1990 Urban et al. \cite{urban}
carried out detailed calculation of this effect proposing this
method as a way to search for antimatter in primary CR at
the TeV energies.

The GMF should deflect the antimatter component in
the CRs in opposite direction with respect to the matter
component. Therefore, if protons are deflected by the GMF towards East, antiprotons are deflected towards West. If the energy is low enough and the angular resolution is adequate we can distinguish, in principle, two shadows, one shifted towards West due to the protons and the other shifted towards East
due to the antiprotons. At high energy ($\geq$ 10 TeV) the
magnetic deflection is too small compared to the angular
resolution and the shadows cannot be disentangled. At low energy
($\approx$100 GeV) the well deflected shadows are washed out by
the poor angular resolution, thus limiting the sensitivity.
Therefore, there is an optimal energy window for the measurement
of the antiproton abundance.

In 1991, the CYGNUS collaboration \cite{cygnus} first observed
the CR shadowing effect measuring a deficit of 4.9
standard deviations (s.d.) in the CR background by superposing the Moon and Sun data at an energy of about 50 TeV. In the same year also the EAS-TOP  experiment observed the shadowing effect due to the Moon and the Sun on the 100 TeV CRs flux with a significance of about 2.7 s.d. \cite{eastop}.
In the following years this effect has been confirmed by other EAS-arrays (CASA-MIA\cite{casamia}, HEGRA \cite{hegra}, GRAPES \cite{grapes}). 

The first observations of a shadowing effect had to wait for the results of the CYGNUS and EAS-TOP experiments in 1991. There are mainly two reasons for this long delay, first the poor angular resolution of EAS-arrays in comparison with the angular radius of the Moon or Sun ($\sim$0.26$^{\circ}$). Indeed, only at the beginning of the 90s the angular resolutions of EAS-arrays reached the 1 deg level. Second, due to the high energy threshold ($\approx$100 TeV) of the experiments the statistical significance of the observations was small and the position of the shadow not affected by the GMF.
Therefore, the deficit of the counting rate was observed as a
function of the angular distance from the Moon position, without any
information on the East-West asymmetry and consequently, without any possibility to study the CR antimatter content.

In 1993, the Tibet AS$\gamma$ experiment measured both the Moon
and Sun shadows with an energy threshold low enough (about 10 TeV)
to allow a 2-dimensional study of the effect. In particular, they
observed for the first time a westward displacement of the Moon
shadow (0.16$^{\circ}$ at the 7.1 $\sigma$ level) from its actual
position. With this result the first upper limit to the
$\overline{p}/p$ ratio with this technique was set at about 30\%
\cite{chantell}. Afterwards, the collaboration set an upper limit to the
$\overline{p}/p$ at 10 TeV at about 10\% \cite{tibet05}.

The CR shadowing effect has been observed also in the high
energy muon distribution with underground detectors (SOUDAN-2
\cite{soudan}, MACRO \cite{macro1,macro2}, L3+C \cite{l3c}, BUST \cite{bust}, MINOS \cite{minos}). Recently, the ICECUBE experiment observed the Moon shadow in the Southern hemisphere with a statistical significance of more than 10 s.d. \cite{icecube11}.
Upper limits to the $\overline{p}/p$ ratio have been set by the MACRO (48\% at 68\% c.l. at about 20 TeV) \cite{macro2} and L3+C (11\%
at 90\% c.l. around 1 TeV) \cite{l3c} collaborations.

High sensitivity observations of the Moon shadow have been
recently reported in the multi-TeV energy region by an upgraded
version of the Tibet AS$\gamma$ array \cite{tibet07} and by the
MILAGRO collaboration \cite{milagro03}. While the Tibet AS$\gamma$
experiment set a limit to the $\overline{p}/p$ ratio to 7\% at
90\% c.l. at about 3 TeV, so far the MILAGRO collaboration did not publish any result on the antimatter search with the Moon shadow.

The ARGO-YBJ experiment is particularly effective in measuring the CR antimatter content via the observation of the CRs shadowing effect due to: (1) good angular resolution, pointing accuracy and long-term stability; (2) low energy threshold; (3) real sensitivity to the GMF due to the absence of any systematic shift in the East-West direction.
In this paper we report the measurement of the $\overline{p}/p$
ratio in the TeV energy region with all the data recorded during the period
from July 2006 through November 2009.

The paper is organized as follows.
In the Section II the ARGO-YBJ detector is described. 
The description of a detailed MC simulation of the Earth-Moon
spectrometer system developed in order to evaluate the deficit of events and to calibrate the detector is briefly sketched out in Section III. 
In Section IV the data analysis is outlined and the detector performance summarized.
The $\bar{p}/p$ ratio calculation method is also described in Section IV.C.
Finally, the results of the data analysis are presented and discussed in Section V. A summary of the obtained results is given in Section VI.

\section{The ARGO-YBJ experiment}

\subsection{The detector}

The ARGO-YBJ experiment, located at the YangBaJing Cosmic Ray
Laboratory (Tibet, P.R. China, 4300 m a.s.l., 606 g/cm$^2$), is constituted by a central carpet $\sim$74$\times$ 78 m$^2$, made of a single layer of Resistive Plate Chambers (RPCs) with $\sim$93$\%$ of active area, enclosed by a guard ring partially instrumented ($\sim$20$\%$) up to $\sim$100$\times$110
m$^2$. 
The RPC is a gaseous detector working with uniform electric field generated by two parallel electrode plates of high bulk resistivity (10$^{11}\Omega$ cm). The intense field of 3.6 kV/mm at 0.6 atm pressure provides very good time resolution (1.8 ns) and the high electrode resistivity limits the area interested by the electrical discharge to few mm$^2$.
The apparatus has a modular structure, the basic data acquisition element being a cluster (5.7$\times$7.6 m$^2$), made of 12 RPCs (2.85$\times$1.23 m$^2$ each). Each chamber is read by 80 external strips of 6.75$\times$61.8 cm$^2$ (the spatial pixel), logically organized in 10 independent pads of 55.6$\times$61.8 cm$^2$ which represent the time pixel of the detector \cite{aielli06}. 
The read-out of 18360 pads and 146880 strips are the experimental output of the detector. 
The RPCs are operated in streamer mode by using a gas mixture (Ar 15\%, Isobutane 10\%, TetraFluoroEthane 75\%) for high altitude operation \cite{bacci00}. The high voltage settled at 7.2 kV ensures an overall efficiency of about 96\% \cite{aielli09a}.
The central carpet contains 130 clusters (hereafter ARGO-130) and the
full detector is composed of 153 clusters for a total active
surface of $\sim$6700 m$^2$. The total instrumented area is $\sim$11000 m$^2$.

A simple, yet powerful, electronic logic has been implemented to build an inclusive trigger. This logic is based on a time correlation between the pad signals depending on their relative distance. In this way, all the shower events giving a number of fired pads N$_{pad}\ge$ N$_{trig}$ in the central carpet in a time window of 420 ns generate the trigger.
This trigger can work with high efficiency down to N$_{trig}$ = 20,
keeping negligible the rate of random coincidences.
The timing calibrations of the pads is performed according to the method reported in \cite{hhh07,aielli09b}.

The whole system, in smooth data taking since July 2006 with ARGO-130, is in stable data taking with the full apparatus of 153 clusters since November 2007 with the trigger condition N$_{trig}$ = 20 and a duty cycle $\geq$85\%. The trigger rate is $\sim$3.5 kHz with a dead time of 4$\%$.

%
Once the coincidence of the secondary particles has been recorded,
the main parameters of the detected shower are reconstructed following the procedure described in \cite{argo-moon11}. 
In short, the reconstruction is split into the following steps. Firstly, the shower core position is derived with the Maximum Likelihood method from the lateral density distribution of the secondary particles. In the second step, given the core position, the shower axis is reconstructed by means of an iterative un-weighted planar fit able to reject the time values belonging to the non-gaussian tails of the arrival time distribution. Finally, a conical correction is applied to the surviving hits in order to improve the angular resolution.
Unlike the information on the plane surface, the conical correction is obtained via a weighted fit which lowers the contribution from delayed secondary particles, not belonging to the shower front. 

The analysis reported in this paper refers to events selected according to
the following criteria: (1) more than 25 strips N$_{strip}$ should be fired on the ARGO-130 carpet; (2) the zenith angle of the shower arrival direction should be less than 50$^{\circ}$; (3) the reconstructed core position should be inside an area 150$\times$150 m$^2$ centered on the detector. 
After these selections the number of events analyzed is about 2.5$\times$10$^{11}$ (about 10$^9$ inside a 10$^{\circ}\times$10$^{\circ}$ angular region centered on the Moon position).
According to simulations, the median energy of the selected protons is E$_{50}\approx$1.8 TeV (mode energy $\approx$0.7 TeV).

\section{Monte Carlo simulation}

The deficit of events in the Moon direction, the absolute energy
calibration and the angular resolution, as well as the systematic
pointing biases, have been studied by comparing the observed Moon
shadow characteristics (East-West and North-South displacements
and shape) with the expectations from a detailed MC simulation of the CR propagation in the Earth-Moon system \cite{argo-moon11,mcmoon,argomoon_icrc11}. 

With this simulation we also estimated the expected antiprotons flux in the opposite CR Moon shadow side, as described in Section IV.C.

From the MC simulation strategy viewpoint, the Moon shadow has been treated like an extensive excess signal, instead of a lack in the isotropic CRs flux. In other words, our simulation deals with the Moon as if it was the source of the CRs which it intercepts in reality.

The simulation has been realized on the basis of the real data acquisition time. The Moon position has been computed at fixed times, starting from July 2006 up to November 2009. Such instants are distant 30 seconds each other. For each time, after checking the data acquisition was effectively running and the Moon was in the field of view, extensive primaries are generated with arrival direction sampled within the Moon disc. For each chemical species (p, He, CNO group, Mg-Si group and Fe), the number of primaries to be generated is computed on the basis of the effective exposure time, according to the energy spectrum resulting from a global fit of the main experimental data \cite{horandel}.

Once the number of CRs expected to be hampered by the Moon has been calculated, the charge sign of every primary is inverted and it is propagated back to the Moon, the magnetic field bending its trajectory. The propagation stops anyway at the Moon distance, giving a good approximation of the deflection undergone by the CR before reaching the atmosphere. In fact, if we firstly consider a positively charged CR arriving to the Earth atmosphere and then we invert its charge and its momentum and threw it back to the space, the two trajectories do not overlap because of the numerical approximation. Nonetheless, as long as the primary energy is above several tenth of GeV, both trajectories give similar deviations, the difference ranging from $15\%$ at $50\textrm{ GeV}$ down to $3\%$ above $1\textrm{ TeV}$. Further details and results from this simulation can be found in \cite{mcmoon,argo-moon11}.

After accounting for the arrival direction correction ought to the magnetic bending effect, the air showers development in the atmosphere has been generated with the CORSIKA v. 6.500 code \cite{corsika}. 
The electromagnetic interactions are described by the EGS4 package while the hadronic interactions above 80 GeV are reproduced by the QGSJET-II.03 and the SYBILL models. The low energy hadronic interactions are described by the FLUKA package. CR spectra have been simulated in the energy range from 10 GeV to 1 PeV following the relative normalization given in \cite{horandel}. About 10$^8$ showers have been sampled in the zenith angle interval 0-60 degrees. The secondary particles have been propagated down to a cut-off energies of 1 MeV (electromagnetic component) and 100 MeV (muons and hadrons).
The experimental conditions (trigger logic, time resolution, electronic noises, relation between strip and pad multiplicity, etc.) have been taken into account via a GEANT4-based code \cite{geant4}. The core positions have been randomly sampled in an energy-dependent area large up to 2$\cdot$10$^3$ $\times$ 2$\cdot$10$^3$ m$^2$, centered on the detector. Simulated events have been generated in the same format used for the experimental data and analyzed with the same reconstruction code.

\section{Data analysis}

For the analysis of the shadowing effect, the signal is collected within a 10$^{\circ}\times$10$^{\circ}$ sky region centered on the Moon position. We used celestial coordinates (right ascension and declination, R.A. and DEC. hereafter) to build the \emph{event} and \emph{background} sky maps, with 0.1$^{\circ}\times$0.1$^{\circ}$ bin size.
Finally, after a smoothing procedure, the \emph{significance} map, used to estimate the statistical significance of the observation, is obtained.

\bfi{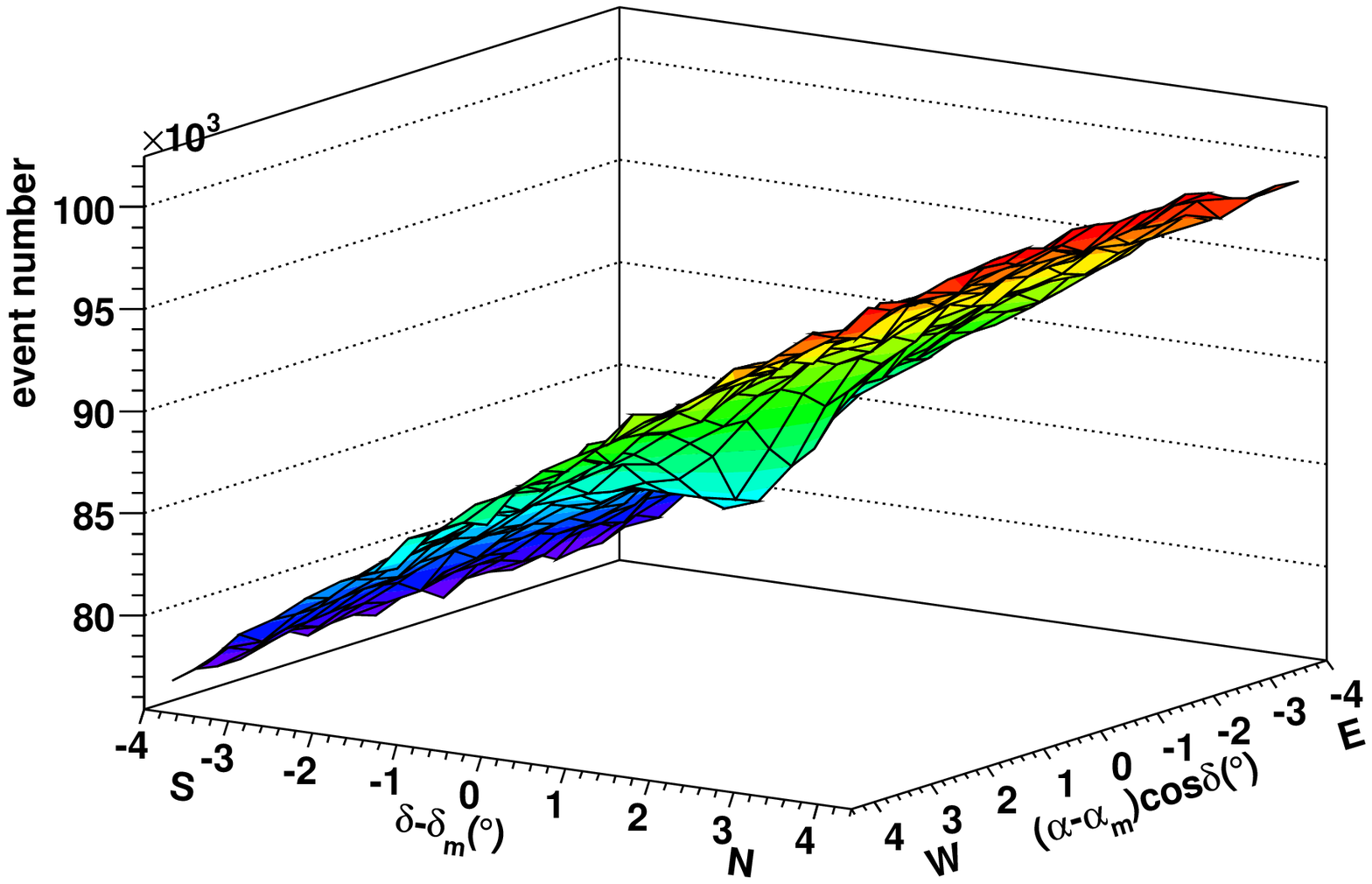}\\
(a)\\
\epsfxsize=9cm \epsffile{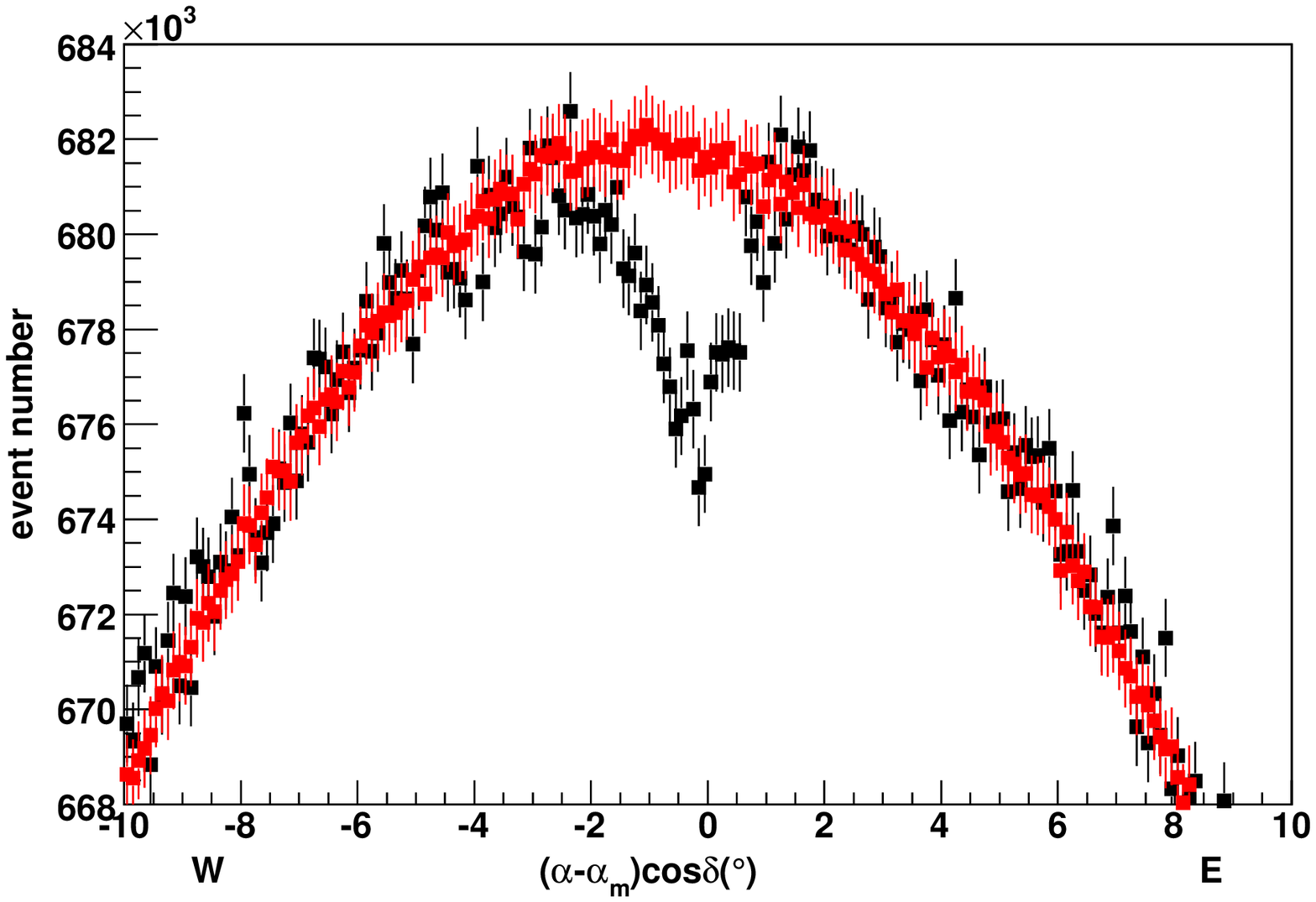}\\
(b)\caption{Plot (a): showers firing N$_{strip}>$100 collected around the Moon position.   The coordinates are R.A. $\alpha$ and DEC. $\delta$ centered on the Moon position ($\alpha_m$, $\delta_m$).
The plot (b) shows the map projections along the R.A. direction.  
\label{fig:moon-evbkg}}
\efi
%
As it can be appreciated in Fig. \ref{fig:moon-evbkg}, the Moon shadow turns out to be a lack in the smooth CR signal, observed by ARGO-YBJ even without subtracting the background contribution nor smoothing the signal. 
\subsection{Moon shadow analysis}

Cosmic rays blocked by the Moon must be as much as the background events lying within a region as large as the Moon disc. A suitable background estimation is therefore a crucial point of the analysis. As it can be seen also from the projection along the R.A. from Fig. \ref{fig:moon-evbkg}, the background events are not uniformly distributed around the Moon, because of the non-uniform exposure of the map bins to the CR radiation.  
The background has been estimated with the \emph{equi-zenith angle} method, as described in detail in \cite{argo-moon11}.

\bfi{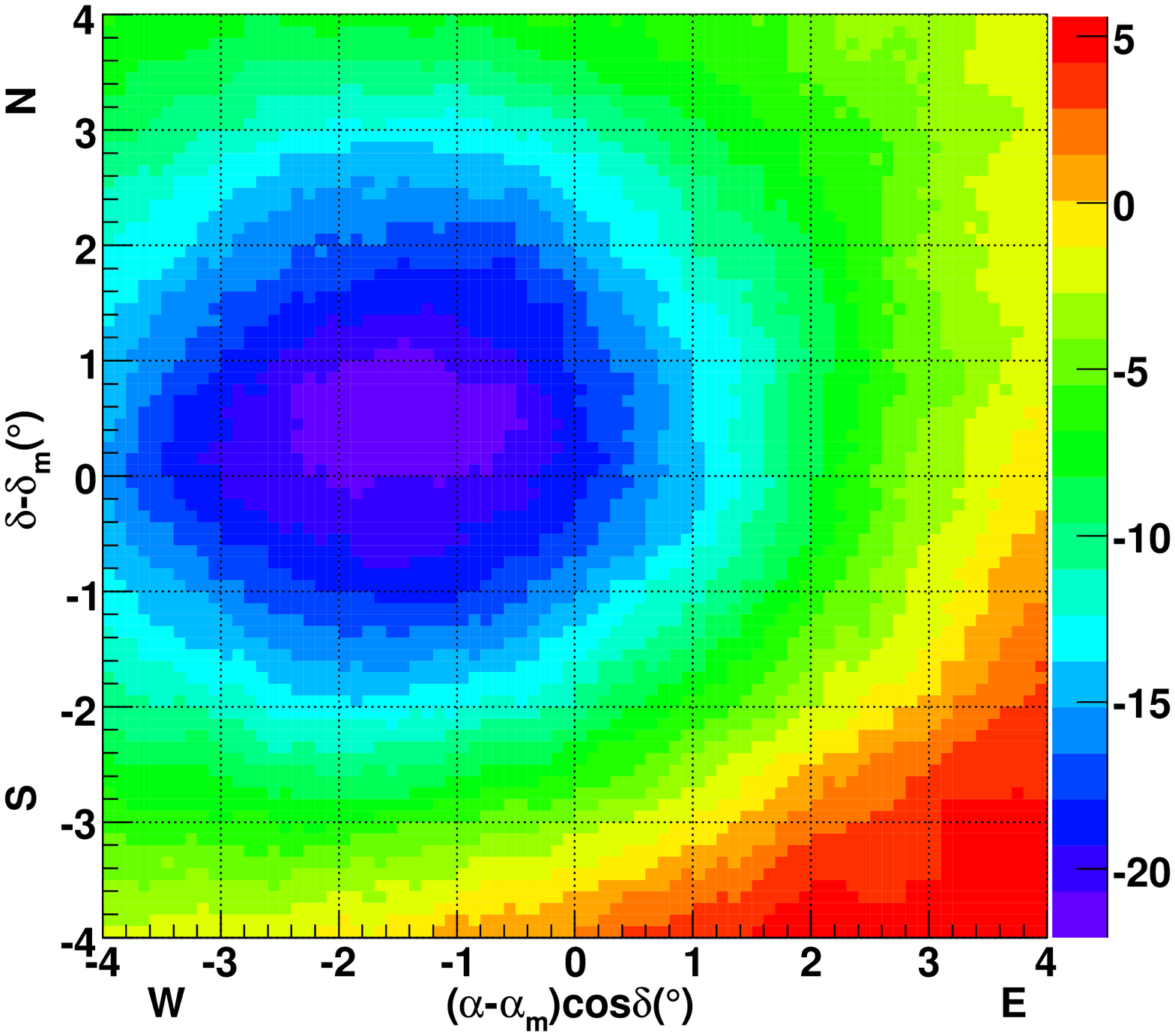}\\
  \caption{Significance map of the Moon region observed with all events detected by ARGO-YBJ. The event multiplicity is 25$\leq N_{strip}<$ 40 and zenith angle $\theta<50^{\circ}$. 
  The coordinates are R.A. $\alpha$ and DEC. $\delta$ centered on the Moon position ($\alpha_m$, $\delta_m$).
  The color scale gives the statistical significance in terms of standard deviations.
 \label{fig:moon1}}
\efi
%
A significance map of the Moon region is shown in Fig. \ref{fig:moon1}. It contains all events belonging to the lowest multiplicity bin investigated (25$\leq N_{strip}<$ 40), collected by ARGO-YBJ during the period July 2006 - November 2009 (about 3200 hours on-source in total). N$_{strip}$ is the number of fired strips on the central carpet ARGO-130.
The significance of the maximum is about 22 s.d..
The observed westward displacement of the Moon shadow by about 1.5$^{\circ}$ allows to appreciate the sensitivity of the ARGO-YBJ experiment to the GMF.
\emph{This means that a potential antiproton signal is expected eastward within 1.5$^{\circ}$ from the actual Moon position} (i.e., within 3$^{\circ}$ from the observed Moon position). 
The median energy of selected events is E$_{50}\approx$750 GeV (mode energy $\approx$ 550 GeV) for proton-induced showers. The corresponding angular resolution is $\sim$1.6$^{\circ}$.

The large displacement of the shadow is only one element of this analysis, the other one being the angular resolution which is not adequate in this multiplicity range. Indeed, as can be seen from the Fig. \ref{fig:moon1}, the matter shadow is visible on the antimatter side with a significance of about 10 s.d., thus limiting the sensitivity to the antiproton abundance measurement.
We note that this is the first time that an EAS array is observing the Moon shadow cast by sub-TeV primary CRs.

\subsection{Detector performance}

The performance of the detector and its operation stability have been studied in detail in \cite{argo-moon11} exploiting the CR Moon shadowing effect with all data since July 2006.

The measured angular resolution is better than 0.5$^{\circ}$ for CR-induced showers with energies E $>$ 5 TeV, in good agreement with MC expectations. 
The Point Spread Function of the detector, studied in the North-South projection not affected by the GMF, is Gaussian for N$_{strip}\geq$200, while for lower multiplicities is better described for both MC and data with a linear combination of two Gaussian functions. The second Gaussian contributes for about 20\%.

The long-term stability of the ARGO-YBJ experiment has been checked by monitoring both the position of the Moon shadow, separately along R.A. and DEC. projections, and the amount of shadow deficit events in the period
November 2007 -- November 2010, for each sidereal month and for events with N$_{strip}>$ 100.
As shown in Fig. 17 of ref. \cite{argo-moon11}, the position of the Moon shadow turned out to be stable at a level of 0.1$^{\circ}$ and the angular resolution stable at a level of 10\%, on a monthly basis. These results make us confident about the detector stability in the long-term observation of the Northern sky.
A systematic uncertainty of (0.19$\pm$0.02)$^{\circ}$ towards the North in the absolute pointing accuracy is observed.  The most important contribution to the
systematics is likely due to a residual effect not completely corrected by the time calibration procedure. Further studies are under way.

We have estimated the primary energy of the detected showers by measuring the westward displacement as a function of the shower multiplicity, thus calibrating the relation between shower size and CR energy. The systematic uncertainty in the absolute rigidity scale is evaluated to be less than 13\% in the range from 1 to 30 TeV/Z, mainly due to the statistical one \cite{argo-moon11}.

\subsection{The CR $\bar{p}/p$ flux ratio calculation}

All chemical species of CR, each with its own spectrum, contribute to form the Moon shadow signal. Hence, the chance of unfolding all contributions relies on MC simulations, as well as the search for antiprotons demands to properly reproduce the Moon shadow signal.
The shape of the Moon shadow is tightly connected to the primaries energy spectrum, which mostly determines the tails of the signal. At first, we assume the energy spectrum of antiprotons follows a power law dN/dE = $k\cdot E^{-\gamma}$, where the spectral index $\gamma$ is taken to be as large as that of protons. An investigation of the dependence on the spectral index will follow in the Section V.

\bfi{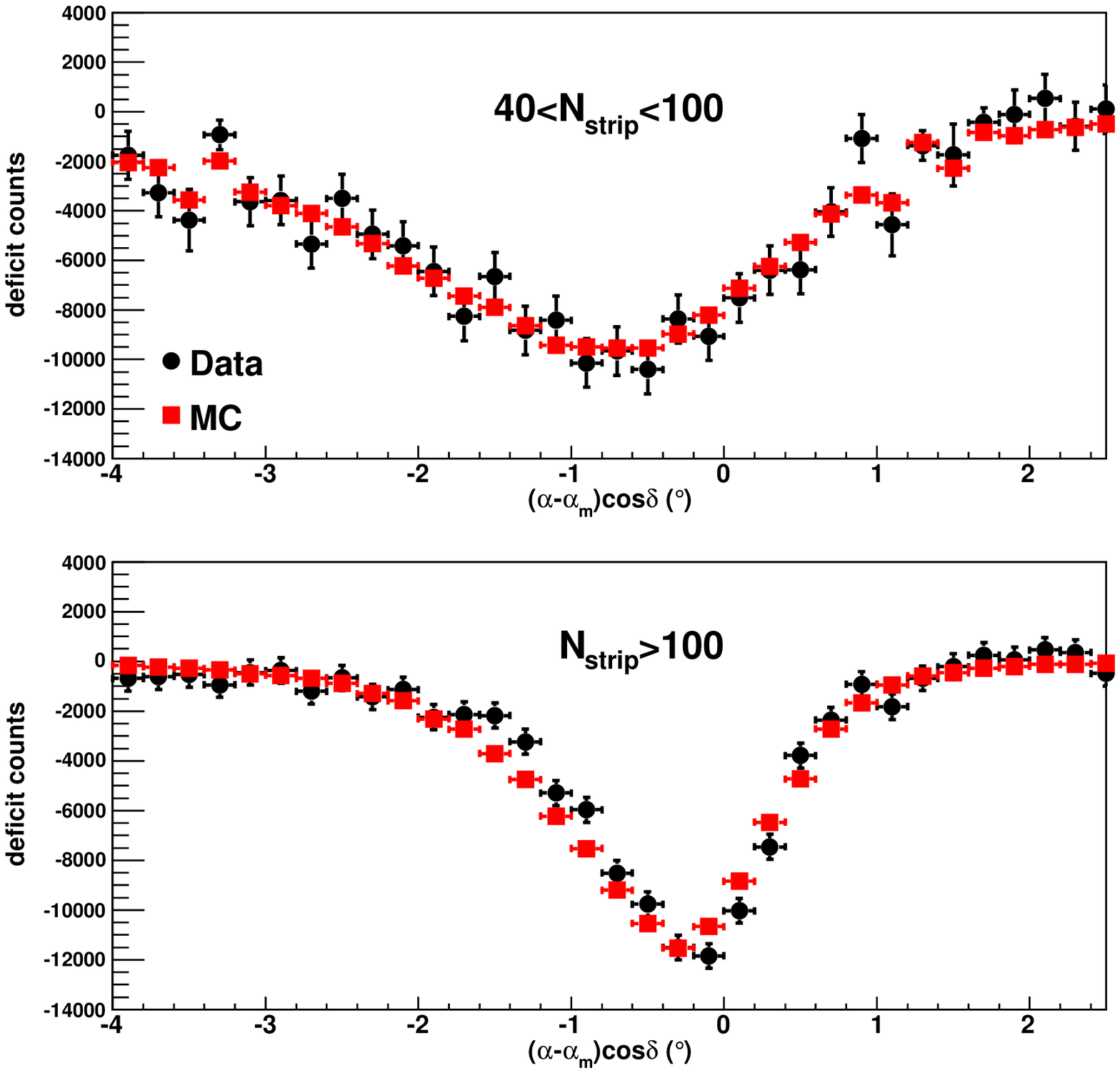}\\
\caption{Deficit counts measured around the Moon projected along
the East-West axis for two different multiplicity bins (black circles) compared to MC expectations (red squares). Events contained in an angular band parallel to the East-West axis and centered on the observed Moon position, proportional to the multiplicity-dependent angular resolution, are used (see text).
  \label{fig:proj-ew}}
\efi
%
In order to evaluate the CR $\bar{p}/p$ flux ratio, only the projection along the R.A. is important, as it has been shown that at Yangbajing the GMF has a non-null effect on the CR trajectories only along such direction \cite{argo-moon11}. The R.A. projection results from an integration along the DEC. direction. 

The deficit counts observed around the Moon projected on the R.A. axis
(that is on the East-West axis) are shown in Fig. \ref{fig:proj-ew} for two multiplicity bins compared to MC expectations: 40$\leq$N$_{strip}<$100 and N$_{strip}>$100, in the upper and lower panels, respectively. 
The vertical axis reports the events contained in an angular band parallel to the East-West axis and centered on the observed Moon position. The widths of these bands are chosen on the basis of the MC simulation so that the shadow deficit is maximized. They turn out to be proportional to the N$_{strip}$-dependent angular resolution. The widths of these bands are
$\pm$2.80$^{\circ}$ and $\pm$1.90$^{\circ}$, respectively. 

The data are in good agreement with the MC simulation and the observed shadows are shifted westward of (-0.75$\pm$0.05)$^{\circ}$ and (-0.30$\pm$0.05)$^{\circ}$, as expected. A detailed analysis of this sort of projections as a function of the shower size is given in \cite{argo-moon11}. 

The GMF shifts westward the dip of the signal from positively charged primaries. Searching antiprotons means looking for excesses in the eastern part of the R.A. projection, i.e. trying and fitting the Moon shape expected from combining CR and antimatter to the shape obtained from experimental data.
Of course, whichever matter-antiprotons combination is obtained, the total amount of triggered events must not be changed, so that the fitting procedure consists in transferring MC events from the CR to the antiprotons shadow and comparing the result with data.
 
To make such a comparison, we firstly adopted the following method. We obtained two kinds of Moon shadow, cast by all CRs and protons, respectively. After projecting them along the R.A. direction, we used a superposition of several Gaussian functions to describe the deficit event distribution in each shadow \cite{tibet07}. Four Gaussian functions were found to be adequate for fitting both distributions within 5$^{\circ}$ from the Moon disc center. Let us name $\theta$ the angular distance from the Moon disc center and $f_m(\theta)$ the Gaussian function superposition describing the CR shadow. Let $\mathcal{F}_{p}(\theta)$ be the proton shadow, obtained by imposing a given power law spectrum. The observed Moon shadow should be expressed by the following function:
%
  \setlength{\arraycolsep}{0.0em}
  \begin{eqnarray*}
f_{MOON}(\theta)=(1-r)\,f_{m}(\theta)+r\,\mathcal{F}_{\overline{p}}(\theta)\\ =(1-r)\,f_{m}(\theta)+r\,\mathcal{F}_{p}(-\theta)
  \end{eqnarray*}
  \setlength{\arraycolsep}{5pt}
%
($0\leq r<1$) where the first term represents the deficit in CRs and the second term represents the deficit in antiprotons. This function must be fitted to the data to obtain the best value of $r$. 

We also applied a second method to determine the antiproton content in the cosmic radiation. Without introducing functions to parameterize the expectations, we directly compared the MC signal with the data. We performed a Maximum Likelihood fit using the $\bar{p}$ content as a free parameter with the following procedure:
\begin{enumerate}
 \item the Moon shadow R.A. projection has been drawn both for data and MC.
 \item the MC Moon shadow has been split into a ``matter'' part
 \emph{plus} an ``antiproton'' part, again so that the
 \emph{total amount of triggered events remains unchanged}:
  \setlength{\arraycolsep}{0.0em}
  \begin{eqnarray*}
\Phi_{MC}(mat) \longrightarrow \Phi_{MC}(r;mat+\bar{p})\\
=(1-r)\Phi_{MC}(mat)+\Phi_{MC}(\bar{p})
  \end{eqnarray*}
  \setlength{\arraycolsep}{5pt}
 \item for each antiproton to matter ratio, the expected Moon shadow
R.A. projection $\Phi_{MC}(r;mat+\bar{p})$ is compared with the
experimental one via the calculation of the likelihood function:
%
$$ \mathcal L (r)=\sum_{i=1}^B N_i ln[E_i(r)]-E_i(r)-ln(N_i!)$$
%
where N$_i$ is the number of experimental events included
within the $i$-th bin, while $E_i(r)$ is the number of
events expected within the same bin, which is calculated by adding the contribution expected from MC ($\Phi_{MC}(r;mat+\bar{p})$) to the \emph{measured} background.
\end{enumerate}

Both methods described above give results consistent within 10\%. 

\section{Results and discussion}

The optimal energy windows for the measurement of the antiproton abundance in CRs is identified by the following multiplicity ranges: 40$\leq$N$_{strip}<$100 and N$_{strip}\geq$100.
In the former bin the statistical significance of Moon shadow observation is 34 s.d., the measured angular resolution is $\sim$1$^{\circ}$, the proton median energy is 1.4 TeV and the number of deficit events about 183000. The shadow is shifted of about 1$^{\circ}$ westward.
In the latter multiplicity bin the significance is 55 s.d., the measured angular resolution $\sim$0.6$^{\circ}$, the proton median energy is 5 TeV and the number of deficit events about 46500. The shadow is shifted of about 0.4$^{\circ}$ westward.

There exists, however, no evidence indicating deficits of CRs at the opposite positions around $\theta$ = 1$^{\circ}$ and 0.4$^{\circ}$ in the eastward direction, corresponding to the particles with negative charge (anti-matter) such as $\bar{p}$, $\bar{He}$, $\bar{C}$,... and $\bar{Fe}$, if any.

Therefore, we applied both methods described in the previous section to evaluate upper limits to the CR $\bar{p}$/p flux ratio.
The $r$ parameter which best fits the expectations to the data turns out to be always negative, i.e. it assumes non-physical values throughout the whole energy range investigated. 
With a direct comparison of the R.A. projections, the $r$-values which maximize the likelihood are: -0.076$\pm$0.040 and -0.144$\pm$0.085 for 40$\leq$N$_{strip}<$100 and N$_{strip}\geq$100, respectively. The corresponding upper limits with 90\% confidence level (c.l.), according to the unified Feldman \& Cousins approach \cite{feldman}, are 0.034 and 0.041, respectively.

\begin{figure*}
\epsfxsize=14cm
\epsffile{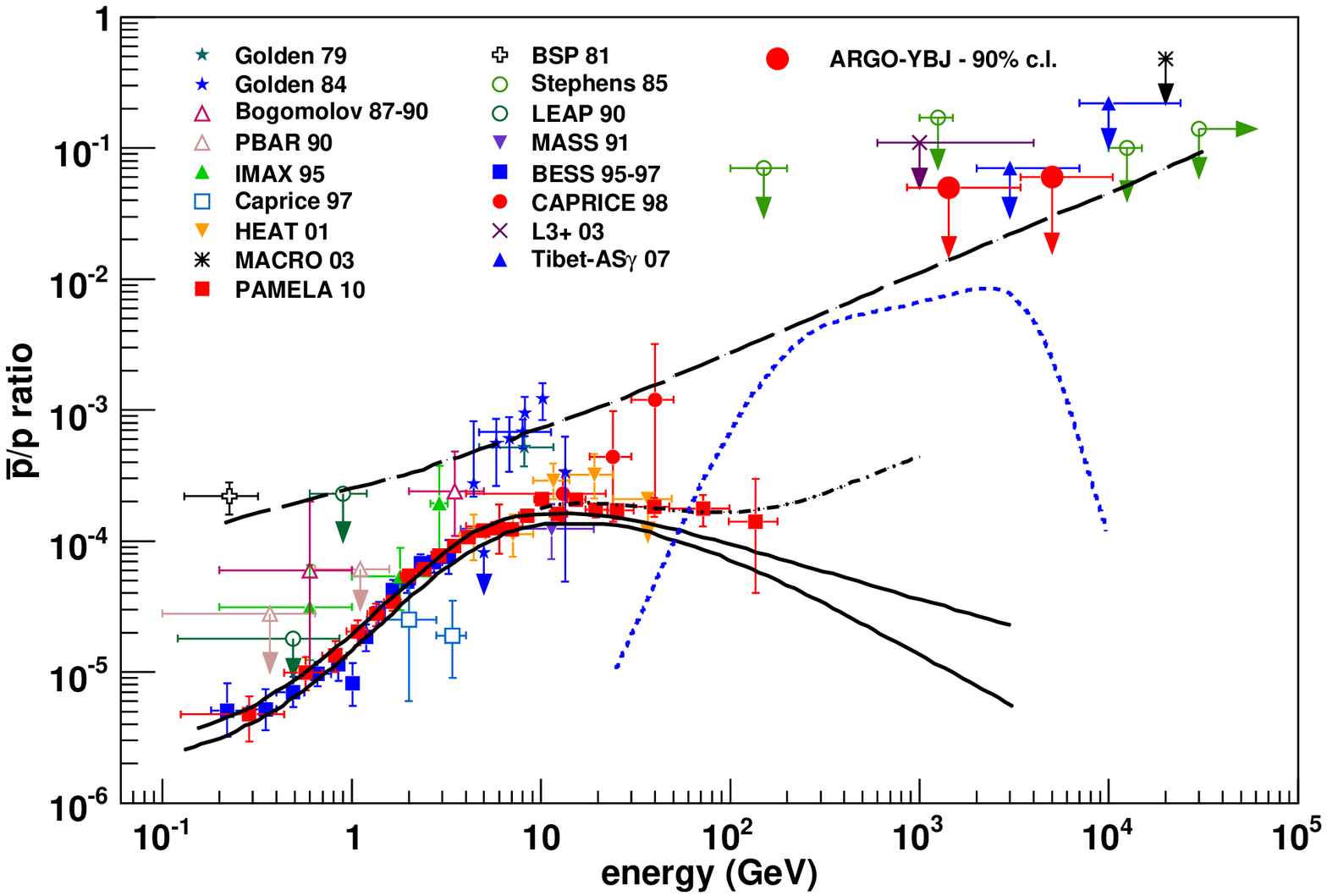}
  \caption{The antiproton to proton flux ratio obtained with the
ARGO-YBJ experiment compared with all the available measurements. 
The solid curve refers to a theoretical calculations for a pure secondary production of antiprotons during the CR propagation in the Galaxy by Donato et al. \cite{donato09}. 
The long-dashed line refer to a model of extragalactic primary $\bar{p}$ production \cite{golden,stephens85}. The rigidity-dependent confinement of CRs in the Galaxy is assumed to be $\propto$ R$^{-\delta}$, where $\delta$ = 0.6. The dotted line refers to the contribution of antiprotons from the annihilation of a heavy dark matter particle \cite{cirelli}.
The short-dashed line shows the calculation by Blasi and Serpico \cite{blasip} for secondary antiprotons including an additional $\bar{p}$ component produced and accelerated at CR sources.
  \label{AntiProton}}
\end{figure*}
%
Since the anti-shadow was assumed to be the mirror image of the proton shadow, we assume for the antiprotons the same median energy. 
The $\bar{p}/p$ ratio is $\Phi(\bar{p})/\Phi(p)$ = 1/$f_p\cdot$ $\Phi(\bar{p})/\Phi(matter)$, therefore, being the assumed proton fraction $f_p$ = 73\% for 40$\leq$N$_{strip}<$100 and $f_p$ = 71\% for N$_{strip}\geq$100 \cite{horandel}, we are able to set the following upper limits at 90\% c.l.: 0.05 for 40$\leq$N$_{strip}<$100 and 0.06 for N$_{strip}\geq$100.
Notice that the two values are similar, in spite of the different multiplicity interval. It is a consequence of the combination of the two opposite effects of the angular resolution and of the geomagnetic deviation.

In Fig. \ref{AntiProton} the ARGO-YBJ results are compared with all the available measurements. The energy bin is 34\% around the median energy for each multiplicity interval. The solid curves refer to a theoretical calculations for a pure secondary production of antiprotons during the CR propagation in the Galaxy by Donato et al. \cite{donato09}. The curves were obtained using the appropriate solar modulation parameter for the PAMELA data taking period \cite{pamelap2}.
The long-dashed line refer to a model of extragalactic primary $\bar{p}$ production \cite{golden,stephens85}. The rigidity-dependent confinement of CRs in the Galaxy is assumed $\propto$ R$^{-\delta}$, being R the rigidity, and $\delta$ = 0.6. 
We note, however, that this curve has been normalized by authors to the low energy $\bar{p}$/p measurements carried out in 1980s. Recent measurements show that the sub-TeV $\bar{p}$/p flux ratio is about a factor 10 lower.
The dotted line refers to a possible contribution of antiprotons from the annihilation of a heavy dark matter particle \cite{cirelli}.
The short-dashed line shows the calculation by Blasi and Serpico \cite{blasip} for secondary antiprotons including an additional $\bar{p}$ component produced and accelerated at CR sources.

Two sources of systematic errors have been investigated: the presence of electrons in the cosmic radiation and the unknown antiproton spectral index. As discussed in \cite{argo-moon11} the absolute rigidity scale uncertainties associated to the CR chemical composition and to different hadronic models in the MC calculations are about 7\% and 12\% respectively.

Like antiprotons, electrons are supposed to shift westward, giving a Moon shadow opposite to that produced by positively charged CR. Nonetheless, the Moon shadow provided by electrons is expected to be slightly different from that of antiprotons. Firstly, for a given multiplicity range the median energy of electron-induced showers is 30\%-40\% less than that of CR showers, i.e. the shadow dip is expected to be further displaced 30\% - 40\%. Then, the angular resolution is better for electron primaries than for CRs ($\sim$30\%). Last, the electron flux at TeV energies is less than 10$^{-3}$ of CR flux \cite{hess}. As a consequence, we estimate that the systematic uncertainty due to misinterpreting electrons as antiprotons is below 10\%.

Since the spectral index of antiprotons is unknown, there is no reason to assume the proton spectral index.
Many unknown factors contribute to its value, mostly related to the diffusion coefficient inside galaxy. 
To investigate this point, primary antiprotons are assigned different spectral indices, $\gamma$= 2.0, 2.2, 2.4, 2.6, 2.8, 3.0. Results for the investigated multiplicity intervals are summarized in table \ref{Table:AntiProton}.
The limits of the antiproton/proton ratio varies of 20\%-30\% with the spectral index.
%
\begin{center}
  \begin{table}[!htb] {\scriptsize \caption{The effect of different spectral indices}
      \begin{tabular}{c|c|c}
        \hline \hline  index & 90\% upper limit(40-100)& 90\% upper
        limit($>$100)\\\hline
        2.0 &     3\%       &      4\%\\
        2.2 &     4\%       &      4\%\\
        2.4 &     4\%       &      4\%\\
        2.6 &     5\%       &      5\%\\
        2.8 &     5\%       &      7\%\\
        3.0 &     6\%       &      7\%\\
        \hline
        \hline
      \end{tabular}
      \label{Table:AntiProton} }
  \end{table}
\end{center}
%

\section{Conclusions}

The ARGO-YBJ experiment is observing the Moon shadow with high statistical significance at an energy threshold of a few hundreds of GeV. Using all data collected until November 2009, we set the upper limits on the $\bar{p}/p$ flux ratio to 5\% at an energy of 1.4 TeV and 6\% at 5 TeV with a confidence level of 90\%. 

In the few-TeV energy range the ARGO-YBJ results provide the strongest $\bar{p}/p$ limits obtained to date, useful to constrain any primary antiproton production model which foresees high fluxes at TeV energies. As discussed in Section IV.A the main limiting factor in the $\bar{p}/p$ ratio measurement exploiting the Moon shadow technique is the angular resolution. The new generation of EAS-arrays under construction (HAWC, LHAASO) is expected to improve the angular resolution by a factor of $\approx$ 3 in the TeV energy range \cite{hawc,lhaaso}. Taking into account also an expected increase of the effective area by at least a factor of 3, we expect that in the next future the sensitivity to the $\bar{p}/p$ ratio could be lowered by a factor of 10 in the TeV energy range.

\begin{acknowledgments}
This work is supported in China by NSFC (No. 10120130794), the Chinese Ministry of Science and Technology,
the Chinese Academy of Sciences, the Key Laboratory
of Particle Astrophysics, CAS, and in Italy by the Istituto
Nazionale di Fisica Nucleare (INFN). We also acknowledge
the essential supports of W. Y. Chen, G. Yang, X. F. Yuan, C.
Y. Zhao, R. Assiro, B. Biondo, S. Bricola, F. Budano, A. Corvaglia,
B. D'Aquino, R. Esposito, A. Innocente, A. Mangano,
E. Pastori, C. Pinto, E. Reali, F. Taurino, and A. Zerbini, in the
installation, debugging, and maintenance of the detector.
\end{acknowledgments}


\end{document}